\documentclass{article}
\usepackage{graphicx} % Required for inserting images
\usepackage{will-math}
\usepackage{graphicx}
\usepackage{subcaption}
\usepackage[labelfont=bf]{caption}
\graphicspath{{./plots/}}
\usepackage[a4paper, total={6in, 8in}]{geometry}
\usepackage{float}
\usepackage[affil-it]{authblk}
\usepackage[superscript,biblabel]{cite}
\usepackage{booktabs}

\newcommand{\vect}[1]{\textbf{#1}}

\title{Determination of unscaled blood input for human dynamic FDG brain PET}
\author[1,2]{Terrell, William}
\affil[1]{Department of Computer Science, University of Virginia}
\affil[2]{Department of Radiology and Medical Imaging, University of Virginia}
\author[3]{Muzi, Mark}
\affil[3]{Department of Radiology, University of Washington, Seattle}
\author[2,4]{Kundu, Bijoy*}
\affil[4]{Department of Biomedical Engineering, University of Virginia}

\affil[*]{\textup{Corresponding Author}}

\date{March 2025}

\begin{document}

\maketitle

\section{Introduction}
Positron Emission Tomography (PET)\cite{phelps_application_1975}, which images the accumulation of a certain injected tracer in a patient’s body, has seen much success in diagnostic radiology\cite{binneboese_correlation_2021}. Traditional, static PET scans, however, only capture the accumulation of tracer in the patient at a single point in time. In reality, the concentration of tracer in a certain tissue changes over time\cite{braune_comparison_2019}–described by the tissue’s time activity curve (TAC). Dynamic PET scans, which consist of a sequence of static PET scans, thus provide a more sensitive method of PET scanning and have seen success in various different disease models\cite{quigg_dynamic_2022}. However, many dynamic PET analysis techniques require the blood input function\cite{croteau_image-derived_2010} of the patient, which measures the concentration of tracer in the patient’s blood vessels over time\cite{seshadri_dynamic_2021}. Acquiring the blood input function non-invasively is difficult, and methods to do so are often specific to a certain tracer. Additionally, such methods often require manual, expert input. In particular, many methods require the segmentation of the internal carotid arteries in humans\cite{kundu_interictal_2024, kundu_dynamic_2024}.

\begin{figure}
    \centering
    \includegraphics[width=0.7\textwidth]{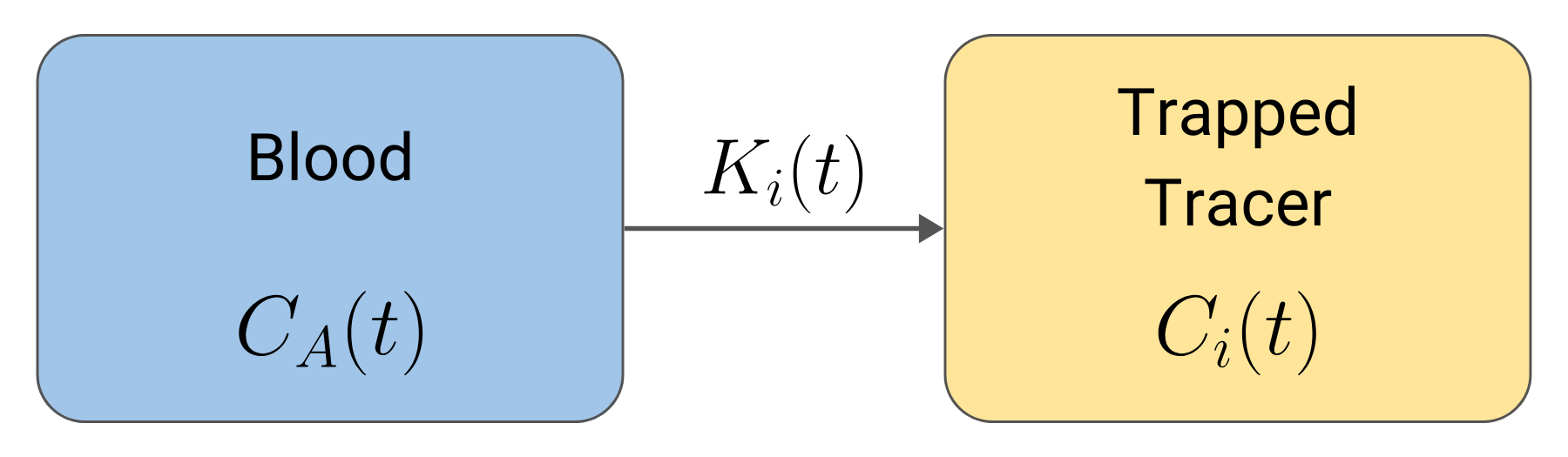}
    \caption{A graphical representation of the Patlak plot. We assume that rate of tracer being trapped is proportional to the amount of blood, which means that $\frac{dC_i}{dt} = K_i C_A(t)$. }
    \label{fig:patlak}
\end{figure}

After obtaining the blood input function, one common and prevalent method for analyzing dynamic PET scans is called the graphical Patlak plot or Patlak method\cite{patlak_graphical_1983}. As shown in Figure \ref{fig:patlak}, the Patlak model is a simple 2-compartment model--we assume that the tracer irreversibly flows from the blood compartment to the tissue compartment at a rate of $K_i$, also known as the rate of uptake. In addition, we assume that the observed concentration of a region is a weighted sum of the concentration in the tissue and the concentration in the blood:
\begin{equation}
    C(t) = C_i(t) + V_b C_A(t) = K_i\int^t_0 C_A(\tau) \, d\tau + V_b C_A(t)
\end{equation}
where $C$ is the observed concentration, $C_A$ is the blood input function, $C_i$ is the tissue uptake, and $V_b$ is another parameter of interest, usually called the volume of blood. If the blood input function $C_A$ is known, one can determine the values of $K_i$ and $V_b$ which best fit each tissue time activity curve and in so doing create a voxel-by-voxel $K_i$ map\cite{seshadri_dynamic_2021, kundu_interictal_2024, kundu_dynamic_2024}. In dynamic FDG-PET, while the Patlak model does not hold from the point of injection, the Patlak model will hold after some time $t^*$, which occurs when certain tracer compartments are in dynamic equilibrium and can be combined into a single compartment\cite{seshadri_dynamic_2021, kundu_interictal_2024, kundu_dynamic_2024}. 

In this paper, we use the key assumption that the Patlak plot holds after $t^*$ to obtain an unscaled version of the blood input function after $t^*$. While this partial blood input function does not suffice for some methods of dynamic analysis, this partial blood input does allow us to use the Patlak model. Using the unscaled blood input function, one can then obtain an unscaled $K_i$ map. In exchange for these tradeoffs, however, our method is fully automated. Additionally, since our method is carefully constructed from first principles, our method is also applicable to every tracer which follows those first principles. Our method does not require any historically obtained ground truth blood input functions either, except for validation.

\section{Methods}

\subsection{Parametrization}

We denote the observed time activity curve for the $n$th voxel of tissue in our image as $C_n(t)$ where $t$ ranges over the time bins $\{t_0, t_1, \dots, t_N\}$ for which we measured the concentration and where $n$ ranges from $1$ to $N_{im}$, the number of voxels in our scan. Practically, $C_n(t)$ will not be a perfect representation of the tissue activity because (i) there is noise present in our observations and (ii) we only know the measured concentration for a finite number of time points. With this in mind, we also denote $C_n^*(t)$ as the true, ideal time activity curve where $t$ now ranges over the interval $[t_0, t_N]$. From this, we can approximate
\begin{equation}
    C_n(t_j) = C_n^*(t_j) + \xi_{n}(t_j)
\end{equation}
for all $j$ and for some unknown noise $\xi_{n}(t_j)$. In reality, $C_n(t_j)$ represents the integral of $C_n^*(t)$ from $t_{j-1}$ to $t_j$; however, this model of $C_n$ is a good enough approximation for our purposes.

We now assume that the Patlak assumption is true for all time points $t$ which are greater than or equal to some $t^* \in [t_0, t_N]$; in other words,
\begin{equation}
    \nonumber
    C^*_n(t) = K_i \int^t_{t_0} C_A(\tau)d\tau + V_b C_A(t)
\end{equation}
for all $t \geq t^*$, for some numbers $K_i$ and $V_b$. Now, let $j^*$ be the first time bin such that $t_{j^*} \geq t^*$. We get that 
\begin{equation}
    \nonumber
    C^*_n(t_j) = K_i \int^{t_j}_{t_0} C_A(\tau)d\tau + V_b C_A(t_j)
\end{equation}
for all $j \geq j^*$. Let $\vect{t} = (t_{j^*}, t_{j^*+1}, ..., t_N)$ and denote $f(\vect{t}) = (f(t_{j^*}), ..., f(t_N))$ for all functions $f : [t_0, t_N] \to \RR^+$. Then, we know that
\begin{equation}
    C^*_n(\vect{t}) = K_i \int^{\vect{t}}_{t_0} C_A(\tau)d\tau + V_b C_A(\vect{t})
\end{equation}
and thus $C^*_n(\vect{t}) \in \text{span}\Big\{\int^{\vect{t}}_{t_0} C_A(\tau)d\tau, C_A(\vect{t})\Big\}$.

Given that the noise in our image, $\xi_{n}(t_j)$, is unbiased for all $j$, we can also say that
\begin{equation}
    \nonumber
    \mathbb{E}[C_n(\vect{t})] = C^*_n(\vect{t}) \in \text{span}\Bigg\{\int^{\vect{t}}_{t_0} C_A(\tau)d\tau, C_A(\vect{t})\Bigg\}.
\end{equation}
Geometrically, this tells us that each of our time activity curves, considered as vectors, lie approximately on the two-dimensional plane which is spanned by the blood input and the integral of the blood input. Therefore, since we have many observed $C_n(\vect{t})$ vectors, we can estimate $\text{span}\Big\{\int^{\vect{t}}_{t_0} C_A(\tau)d\tau, C_A(\vect{t})\Big\}$ by finding the plane which bests fits our vectors $C_n(\vect{t})$. 

\begin{figure}[H]
    \centering
    \includegraphics[width=0.8\linewidth]{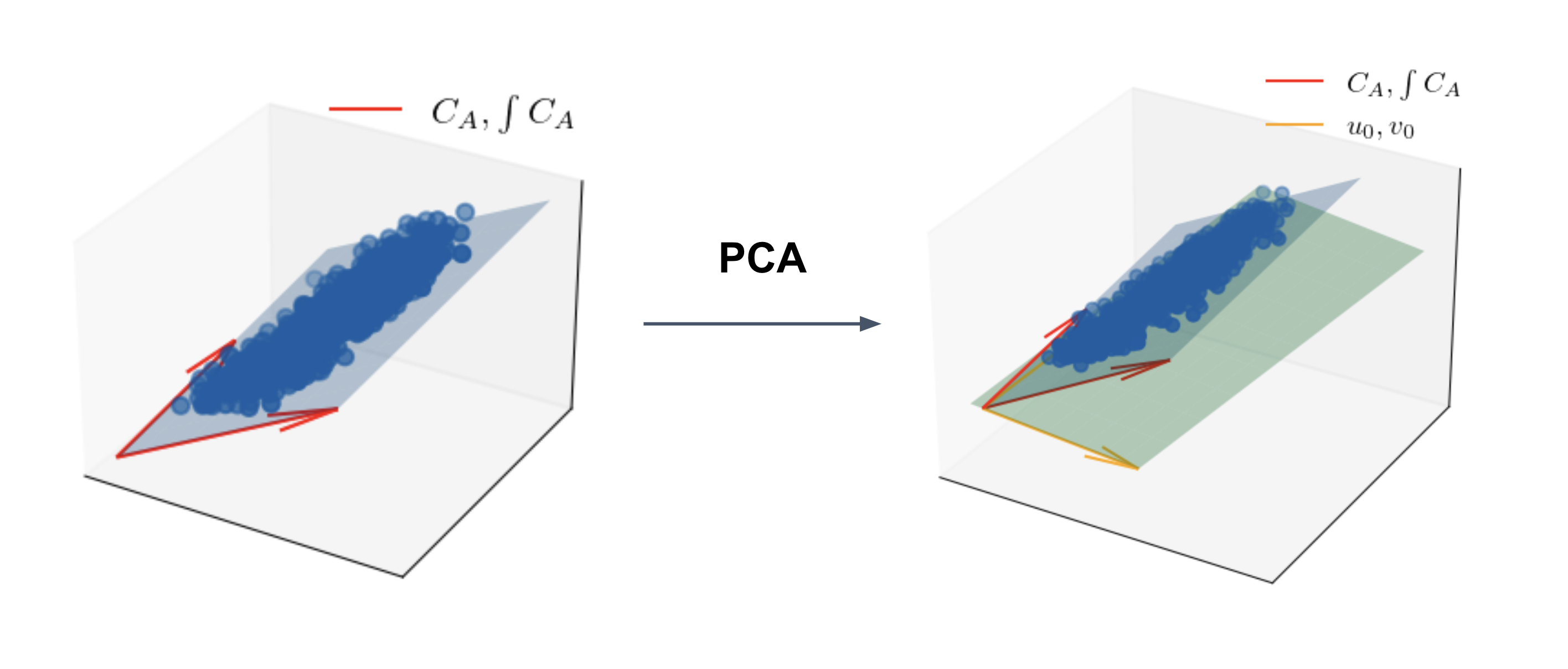}
    \caption{A visualization of how the plane spanned by the blood input function and the integral of the blood input function (red vectors and blue plane) compares to the plane spanned by the two principal vectors obtained from PCA (yellow vectors and green plane) for patient Normal1. A randomized selection of time activity curves (blue points) are displayed as well. To visualize this data, we project down to three dimensions by taking the first coordinate as $C_n(t_{j^*})$, the second as $C_n(t_{j^*+1})+C_n(t_{j^*+2})+C_n(t_{j^*+3})$, and the third as $\sum^{N}_{i=j^*+4} C_n(t_{i})$.}
    \label{fig:pca_fit}
\end{figure}

Practically, we can use principal component analysis (PCA)\cite{greenacre_principal_2022} to find two unit length, orthogonal vectors $\vect{u}_0$ and $\vect{v}_0$ which minimize the average sum of square errors between $C_n(\vect{t})$ and its projection onto the plane $\text{span}\{\vect{u}_0, \vect{v}_0\}$. Proceeding, we will assume that our estimate is correct:
\begin{equation}
    P := \text{span}\Bigg\{\int^{\vect{t}}_{t_0} C_A(\tau)d\tau, C_A(\vect{t})\Bigg\} = \text{span}\{\vect{u}_0, \vect{v}_0\}.
\end{equation}
Figure \ref{fig:pca_fit} shows, for a single patient, how well the vector space formed by the first two principal components of the set of tissue time activity curves compares to the vector space formed by the blood and integral of the blood and how well both match the underlying data.

The importance of this observation is that this plane allows us to parameterize the blood input function and the integral of the blood input function with very few parameters. In fact, to capture the shape of the blood input function without the scale, only one parameter is required. 

\subsection{Optimization}

Now, we wish to derive the blood input function from the basis vectors $\vect{u}_0$ and $\vect{v}_0$. From our previous assumption, we know that $\int^{\vect{t}}_{t_0} C_A(\tau)d\tau \in P$ and $C_A(\vect{t}) \in P$. Additionally, we know that the vectors $\int^{\vect{t}}_{t_0} C_A(\tau)d\tau$ and $C_A(\vect{t})$ satisfy a unique property; namely, that $\int^t_{t_0} C_A(\tau)d\tau$ is the anti-derivative of $C_A(t)$. Therefore, if we have two vectors $\vect{u}=C_i^*(\vect{t}), \vect{v}=C_j^*(\vect{t}) \in P$ for which we believe $\vect{u}$ is the blood input and $\vect{v}$ is the integral of the blood input, 
\begin{equation*}
    \Bigg|\int_{t_0}^\vect{t} \vect{u} - \vect{v}\Bigg| = \Bigg|\int_{t_{j^*}}^\vect{t} \vect{u} + C - \vect{v}\Bigg|
\end{equation*}
must be small in comparison to $|\vect{u}|$ and $|\vect{v}|$ where $\int_a^b \vect{u}$ is a numerical estimate for the integral of $\vect{u}$ from $a$ to $b$ and $C$ is some constant. This prompts us to solve an optimization problem of the form
\begin{equation}
    \underset{C, \vect{u}, |\vect{v}|=1}{\text{min}} \Bigg|\int \vect{u} + C - \vect{v}\Bigg|
    \label{eq:min}
\end{equation}
where $\int \vect{u}$ is now an estimate for the anti-derivative of $\vect{u}$. Note that we must constrain $|\vect{v}| = 1$ since we want the error to be small relative to $|\vect{v}|$; otherwise, we could just continue to decrease the magnitudes of $\vect{u}$ and $\vect{v}$ to get progressively smaller values of the objective function. This property also means that this minimization problem cannot distinguish the magnitude of the estimated blood input function, only the overall shape. Additionally, in practice, we are not necessarily guaranteed to get a global minima $(\vect{u}, \vect{v})$ which matches the blood input function and the integral of the blood input function from this minimization problem. However, given that the anti-derivative of $\vect{u}$ matching with $\vect{v}$ is a necessary condition for $\vect{u}$ and $\vect{v}$ to be the blood and integral of blood, we believe that it is reasonable to assume that there is at least a locally minimal pair of vectors $(\vect{u}, \vect{v})$ which match well with the blood input and the integral of the blood input. We continue under this assumption.

Given a vector $\vect{u} \in P$, we choose to approximate the anti-derivative of $\vect{u}$ by first interpolating on $[t_{j^*}, t_N]$ and then anti-differentiating the resulting interpolation analytically. Therefore, we first construct a model for tissue activity, to use for interpolation. The model
\begin{equation}
    C_A(t) = (A_1(t - \tau) - A_2 - A_3)e^{\lambda_1(t-\tau)} + A_2 e^{\lambda_2(t-\tau)} + A_3 e^{\lambda_3(t-\tau)}
\end{equation}
is commonly used to de-noise the blood input function\cite{feng_models_1993}. For FDG-PET imaging, $\lambda_1$ is negative enough that we can assume the first term is zero after $t^*$. It is also typically the case that $\tau$, the time delay, is far less than $t^*$. Therefore, we approximate
\begin{equation}
    C_A(t) \approx A_2 e^{\lambda_2 t} + A_3 e^{\lambda_3 t}.
\end{equation}
As $P$ is spanned by $C_A(t)$ and its integral, we get that 
\begin{align}
    \nonumber
    u(t) &= A_\text{blood} C_A(t) + A_\text{tissue}\int_0^t C_A(t) \, dt\\
    \nonumber
    &\approx A_\text{blood}\big(A_2 e^{\lambda_2 t} + A_3 e^{\lambda_3 t}\big) +  A_\text{tissue}\bigg(\frac{A_2}{\lambda_2} e^{\lambda_2 t} + \frac{A_3}{\lambda_3} e^{\lambda_3 t} - A_2 - A_3\bigg) \\
    \label{eq:model}
    &=: B_1 e^{\beta_1 t} + B_2 e^{\beta_2 t} + B_3 =: C^*(t; \theta)
\end{align}
where $\theta$ is a vector of parameters and $u(t)$ denotes the true tissue activity function underlying $\vect{u}$. Now, given $\vect{u} \in P$, we solve the minimization problem
\begin{equation}
    \underset{\theta}{\text{min}} |\vect{u} - C^*(\vect{t}; \theta)|
\end{equation}
to interpolate $\vect{u}$. Since we want to interpolate many different choices for $\vect{u}$, we wish to avoid re-solving this minimization every time. Instead, we find $\theta_u$ and $\theta_v$ such that $u_0(\vect{t}) := C^*(\vect{t}; \theta_u) \approx \vect{u}_0$ and $v_0(\vect{t}) := C^*(\vect{t}; \theta_v) \approx \vect{v}_0$. Then, given that $\vect{u} = a_1 \vect{u}_0 + a_2 \vect{v}_0$, we will let $u(t) = a_1 u_0(t) + a_2 v_0(t)$. Note that unless the $\beta_1$ and the $\beta_2$ parameters are the same for both $u_0(t)$ and $v_0(t)$, the $u(t)$ we have obtained will not necessarily be the best-fitting approximation for $\vect{u}$ nor will it even follow the model $C^*(t; \theta)$. In practice, however, these concerns are outweighed by the benefits in speed. If one desires, they can also force the $\beta_1$ and $\beta_2$ parameters to be the same for $u_0(t)$ and $v_0(t)$, which will ensure that $u(t)$ is optimal in the vector space of functions $\text{span}\{1, e^{\beta_1 t}, e^{\beta_2 t}\}$, since the problem reduces to a projection.

Given $u(t) = B_1 e^{\beta_1 t} + B_2 e^{\beta_2 t} + B_3$ on the interval $[t_{j^*}, t_N]$, we get that
\begin{equation}
    \nonumber
    \int^{t}_{t_{j^*}}u(\tau) \, d\tau = \frac{B_1}{\beta_1} e^{\beta_1 t} + \frac{B_2}{\beta_2} e^{\beta_2 t} + B_3 t - B_1 e^{\beta_1 t_{j^*}} - B_2 e^{\beta_2 t_{j^*}}.
\end{equation}
Since we will be adding back a constant in our minimization problem anyway, we can remove the constant terms from this expression and set
\begin{equation}
    \int \vect{u} := \frac{B_1}{\beta_1} e^{\beta_1 \vect{t}} + \frac{B_2}{\beta_2} e^{\beta_2 \vect{t}} + B_3 \vect{t}.
\end{equation}

From here, we have all we need to solve our original minimization problem in equation \ref{eq:min}. However, we can make two extra modifications to reduce the dimensionality of our problem and to bring our problem more in line with a maximum likelihood optimization. 

First, we simplify our optimization by restricting $|\vect{u}| = 1$ in addition to $|\vect{v}|=1$, introducing an additional constant $D$ to optimize, and nesting our optimization:
\begin{equation}
    \underset{|\vect{u}| = |\vect{v}| = 0}{\text{min}} \ \underset{D, C}{\text{min}} \, \bigg|D \int \vect{u} + C - \vect{v} \bigg|.
\end{equation}
The reason we choose to nest the minimization problem is that the inside minimization can be solved exactly via projection:
\begin{equation}
    \nonumber
    \underset{D, C}{\text{min}} \bigg|D \int \vect{u} + C - \vect{v} \bigg| = \Big| \vect{v} - \underset{V}{\text{proj}}  \, \vect{v} \Big|
\end{equation}
where $V = \text{span}\Big\{\int \vect{u}, \vect{1} \Big\}$. In the end, we get the minimization problem
\begin{equation}
    \underset{|\vect{u}| = |\vect{v}| = 0}{\text{min}} \Big| \vect{v} - \underset{V}{\text{proj}}  \, \vect{v} \Big|.
\end{equation}
This loss function only requires two parameters since we can parameterize $\vect{u}$ and $\vect{v}$ like
\begin{align}
    \vect{u}(\theta) &= \text{cos}(\theta) \, \vect{u}_0 + \text{sin}(\theta) \, \vect{v}_0 \\
    \vect{v}(\varphi) &= \text{cos}(\varphi) \, \vect{u}_0 + \text{sin}(\varphi) \, \vect{v}_0 
\end{align}
due to $\vect{u}_0$ and $\vect{v}_0$ being orthogonal, unit length vectors.

Next, rather than assuming that PCA perfectly capturing the plane defined by the blood and integral of blood, assume now that 
\begin{align}
    \nonumber
    \tilde{\vect{u}}_0 &= \vect{u}_0 + \xi_1 \\
    \nonumber
    \tilde{\vect{v}}_0 &= \vect{v}_0 + \xi_2
\end{align}
are the vectors we actually get from PCA while $\xi_1$ and $\xi_2$ are vectors distributed isotropic normal with mean zero and with covariance matrices $\Sigma_1 = \sigma_1^2 I$ and $\Sigma_2 = \sigma_2^2 I$ respectively. Here, note that $\vect{u}_0$ and $\vect{v}_0$ still represent vectors which perfectly capture the plane spanned by the blood and the integral of the blood. Then, given that we know what $\vect{w} = c_1\vect{u}_0 + c_2 \vect{v}_0$ is for some coordinates $(c_1, c_2)$, we want to find the probability of observing a $\tilde{\vect{w}} = c_1\tilde{\vect{u}}_0 + c_2 \tilde{\vect{v}}_0$. Note that $\tilde{\vect{w}} - \vect{w} = c_1\xi_1 + c_2 \xi_2$. As a result, $\tilde{\vect{w}} - \vect{w}$ is distributed $\mathcal{N}\big(\vect{0}, \sigma^2I\big)$ where $\sigma^2 = c_1^2\sigma_1^2 + c_2^2 \sigma_2^2$. Therefore, $|\tilde{\vect{w}} - \vect{w}|/\sigma$ is distributed $\chi^2$.

For our use case, we consider $v(t)$ and $\text{proj}_V \, \vect{v}$ to be noiseless, and we wish to capture the distance between $\vect{v}$ and $\text{proj}_V \, \vect{v}$. If we assume that $\mathbb{E}[\vect{v}] = v(\vect{t})$, we can first estimate the variance by
\begin{equation}
    \hat{\sigma}^2 = \frac{|\vect{v} - v(\vect{t})|^2}{n-1}
\end{equation}
We therefore know that $|\vect{v} - v(\vect{t})|$ is proportional to an estimator of the standard deviation. This means, as with previously, we can change our problem to
\begin{equation}
    \underset{|\vect{u}|=|\vect{v}|=1}{\text{min}} \frac{\big|\vect{v} - \text{proj}_V  \, \vect{v} \big|}{|\vect{v} - v(\vect{t})|}
\end{equation}
to account for some degree of noise in our data. While we believe that this objective function likely isn't optimal and requires more statistical work, this objective function nonetheless worked well for us.

\subsection{Application to Dynamic FDG-PET Imaging}

In dynamic FDG-PET imaging, the Patlak model does describe tissue activity well after some time point $t^* \in [t_0, t_N]$, as assumed previously. However, due to the high amount of noise present in dynamic FDG-PET imaging, applying our methods without alteration produces poor results. In particular, noise greatly affects the identification of the second principal component of tissue activity when we truncate to time points after $t^*$. For this reason, we assume in this paper that we can identify the principal two components of tissue activity after $t^*$ well by first applying PCA to the \textit{full-time} dynamic data and then truncating to time points after $t^*$. This assumption helps greatly with the identification of the second component since the signal before $t^*$ is somewhat more heterogeneous than the signal after $t^*$. More specifically, the second component of tissue activity typically has features consistent with the blood input function---namely a high, early peak and a dropping off tail. Therefore, applying PCA over a time range which captures the early peaks of certain tissue activity functions helps distinguish between the first and second components.

% One can find specific psuedocode for this algorithm in Algorithm 1.

Additionally, we bound our optimization. We first bound our optimization to ensure that the curves we parameterize are nowhere negative. Then, we also bound the optimization so that the estimated blood input function---which, recall, is a weighted sum of the first two principal components of tissue uptake---has a greater weight for the second component than the first component. We bounded our optimization this way since, again, we found that the second component has features which are characteristic of the blood input function.

\section{Results}

We applied our results to twelve normal patients, imaged at the University of Washington. During scanning, 25 arterial blood samples were collected for each patient. The ground truth tracer concentration in the blood was determined from these samples\cite{graham_fdg_2002}, and the resulting time activity curve was de-noised using the 7-parameter model\cite{feng_models_1993} previously mentioned.  We determined that the Patlak assumption was satisfied in each patient’s brain after a $t^*$ of 11.0 minutes, and computed each patient’s ground truth $K_i$ map accordingly. 

Applying our methods, we sought to answer five questions:
\begin{itemize}
    \item \textbf{Q1:} Does the Patlak plot indeed hold for our data? Do the blood input and the integral of the blood input agree with the first two components of tissue uptake after $t^*$?
    \item \textbf{Q2:} Does our model of tissue uptake describe the data well?
    \item \textbf{Q3:} Is our optimization problem well-behaved? Do we require bounding or multiple sets of initial parameters?
    \item \textbf{Q4:} Does the estimated blood input match well with the ground truth blood input?
    \item \textbf{Q5:} Do the estimated and ground truth $K_i$-maps match well?
\end{itemize}

\subsection{Q1: Does the Patlak plot indeed hold for our data? Do the blood input and the integral of the blood input agree with the first two components of tissue uptake after $t^*$?}
\begin{figure}[H]
    \centering
    \includegraphics[width=\textwidth]{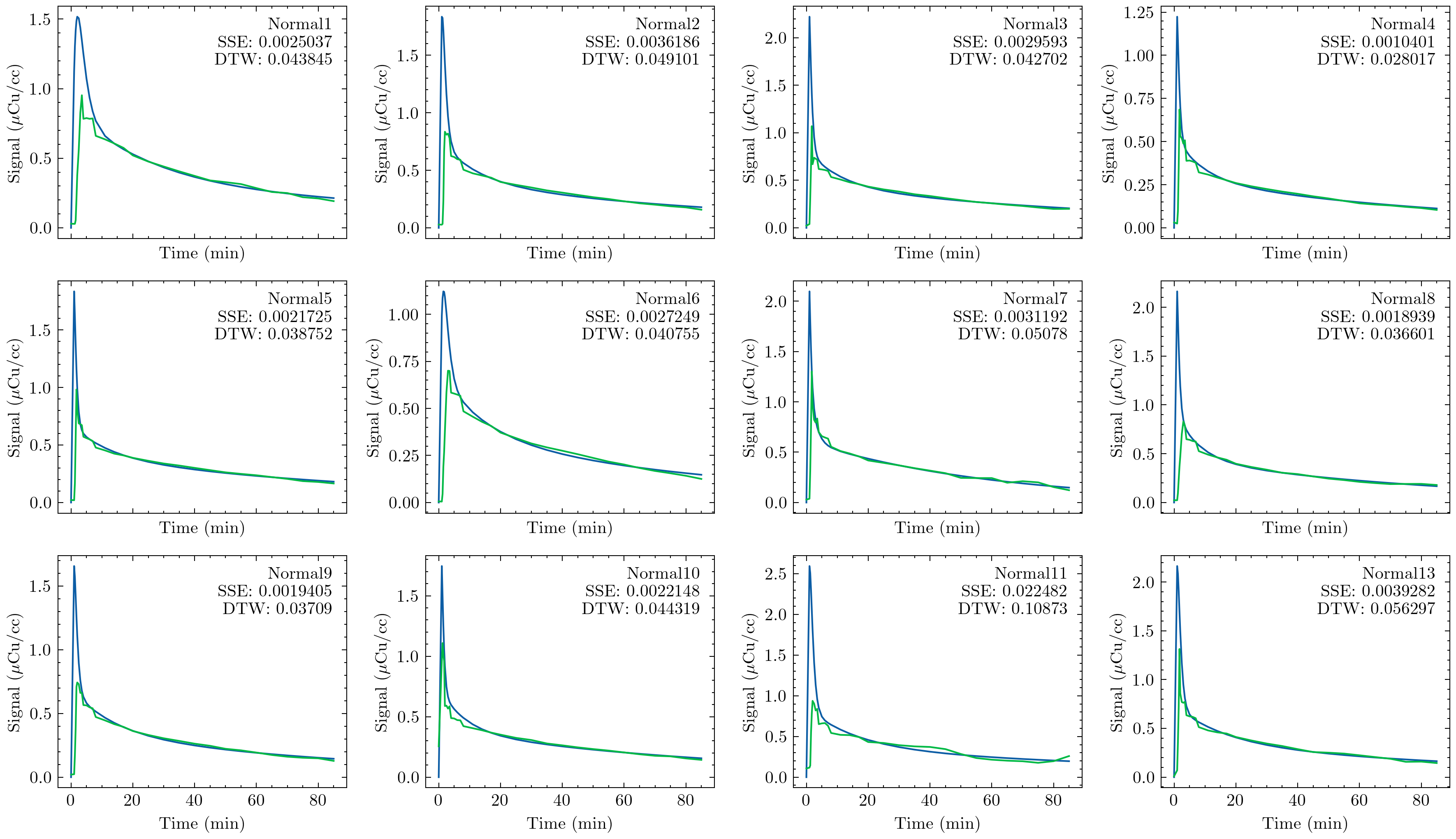}
    \caption{The blood input (blue) versus the linear combination of the first two principal components of tissue uptake which best fits the blood input after $t^*$ (green). The sum of squared errors (SSE) and dynamic time warping distance (DTW) for each patient are shown in the top right. }
    \label{fig:blood_vs_proj}
\end{figure}
Recall that for dynamic FDG-PET images, we assume that (1) the Patlak plot holds after some time $t^*$ and that (2) we can estimate the first two principal components of late-time tissue uptake---which we would conventionally obtain by applying PCA to the set of tissue uptake functions truncated to $[t^*, t_N]$---by applying PCA to the full-time dynamic data. The second of these two assumptions is particularly important to test since the Patlak model does not generally hold for data points before $t^*$, meaning that the full-time data is not guaranteed to be well approximated by a low-dimensional vector space as provided by PCA. 

For the purpose of testing these two assumption simultaneously, we applied PCA to the full-time dynamic data and extracted the principal two components. We then truncated these two components and the blood and integral of blood to the time range $[t^*, t_N]$. Lastly, we found the linear combinations of the truncated principal components which best fit the truncated blood and integral of blood curves via projection. If the projected and ground truth curves match, this indicates that the planes spanned by both sets of curves match.

\begin{figure}[H]
    \centering
    \includegraphics[width=\textwidth]{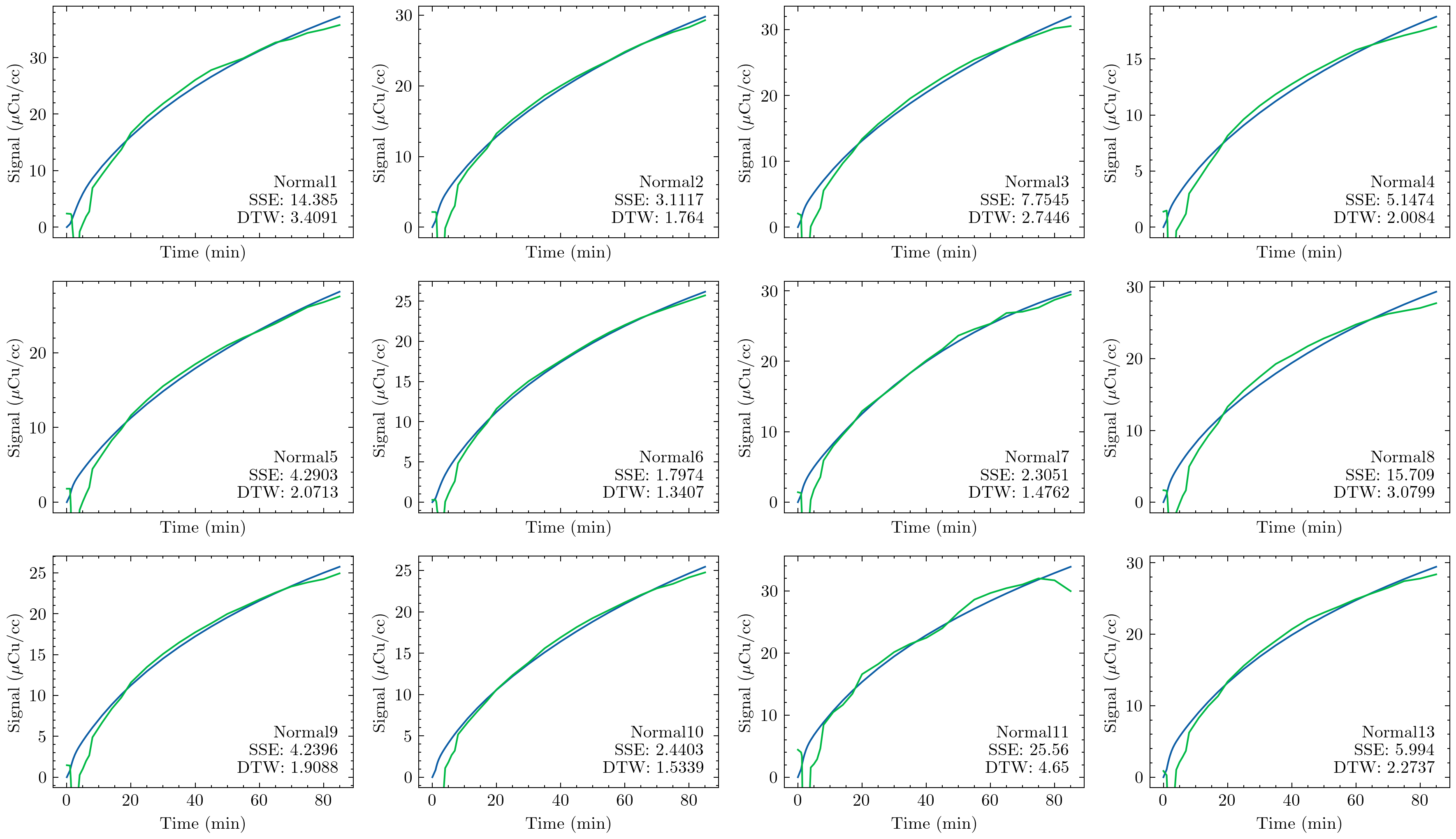}
    \caption{The integral of the blood input (blue) versus the linear combination of the first two principal components of tissue uptake which best fits the integral of the blood after $t^*$ (green). The sum of squared errors (SSE) and dynamic time warping distance (DTW) for each patient are shown in the bottom right.}
    \label{fig:int_vs_proj}
\end{figure}

Figures \ref{fig:blood_vs_proj} and \ref{fig:int_vs_proj} indeed show that there is good correspondence between the blood input and the integral of blood and their projections onto the principal two components of tissue uptake. In fact, we obtained an average sum of squared errors (SSE) of $0.0042165 \pm  0.0055592$ and $7.7279 \pm  6.9132$ between the blood and projected blood and the integral and projected integral respectively. We also obtained an average dynamic time warping distance (DTW) of $0.048083 \pm  0.019604$ and $2.3551 \pm  0.92586$ for the blood and integral of blood respectively. Given that our method parameterizes the estimated blood input function by a vector on the plane spanned by these two principal components, observe that these results give us an upper limit for the performance of our method.

% The scale together with the numbers might be confusing for readers--the sse and the dtw are based off of the original scale of the curves while the images show different scales

\subsection{Q2: Does our model of tissue uptake describe the data well?}
\begin{figure}[H]
    \centering
    \includegraphics[width=\textwidth]{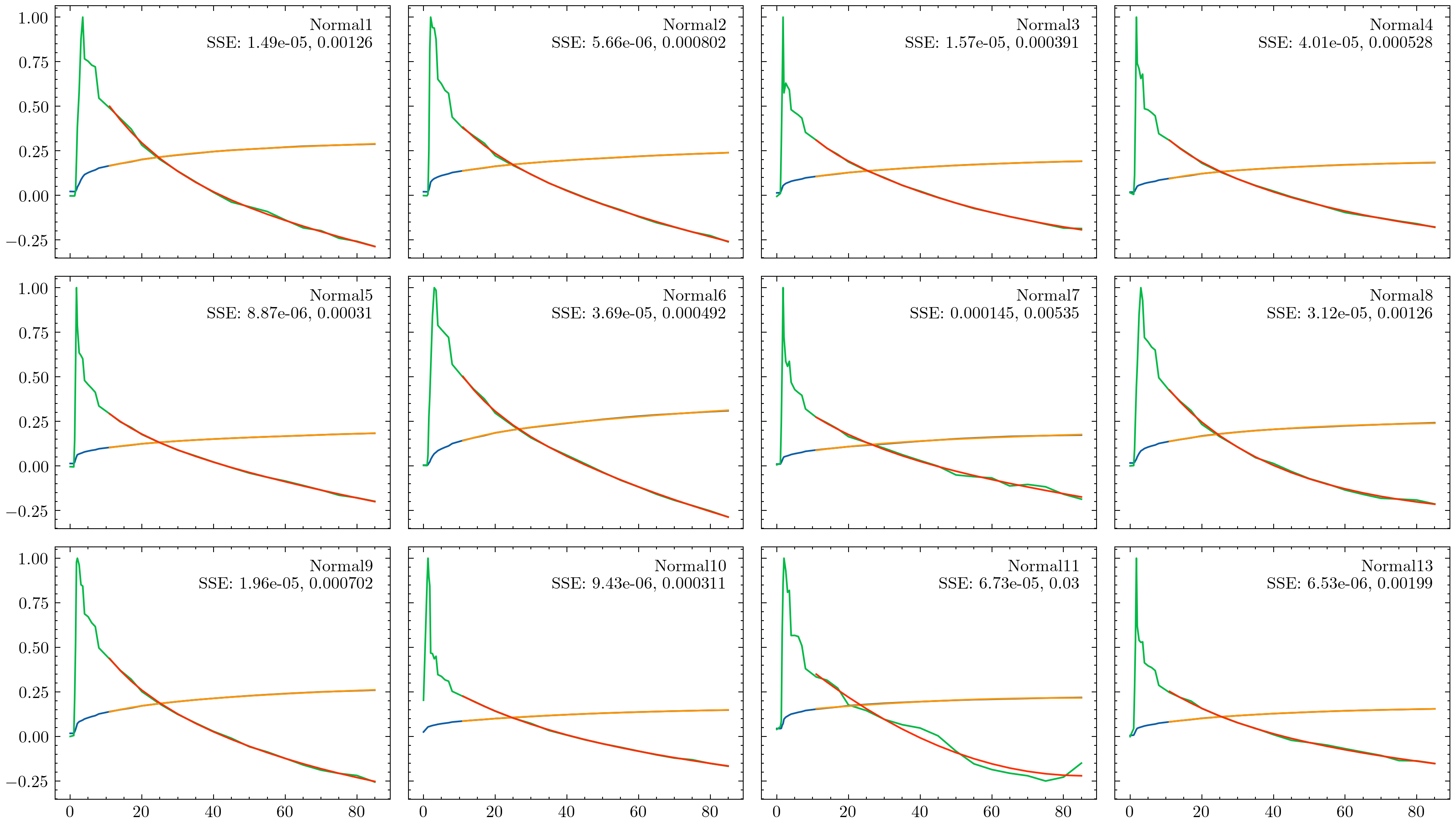}
    \caption{The principal two components of tissue uptake for the full-time dynamic scan versus the curves following our model which best fit these two components at late times points. Blue and green are respectively the first and second principal components while orange and red are their best fitting interpolations.}
    \label{fig:model}
\end{figure}

For our method to work, we also assume that the model
\begin{equation}
    \tag{\ref{eq:model}}
    C^*(t; \theta) = B_1 e^{\beta_1 t} + B_2 e^{\beta_2 t} + B_3
\end{equation}
describes the tissue uptake curves of our dynamic FDG-PET data well. Specifically, we want this model to fit the first two principal components of tissue uptake well. Therefore, for each of our dynamic PET scans, we extracted the first two principal components of tissue uptake and fit this model to time points after $t^*$ using SciPy's \texttt{minimize} with the SLSQP method\cite{dieter_software_1988, 2020SciPy-NMeth}. Ultimately, we obtained an average SSE of $3.3399\mathrm{e}{-5} \pm  3.7743\mathrm{e}{-5}$ and $0.0036152 \pm  0.0080632$ for the first two components respectively and an average DTW of $0.0050834 \pm  0.0026224$ and $0.039342 \pm  0.031225$. Figure \ref{fig:model} shows these results.

\subsection{Q3: Is our optimization problem well-behaved? Do we require bounding or multiple sets of initial parameters?}
\begin{figure}[H]
    \centering
    \begin{subfigure}{.5\textwidth}
        \centering
        \includegraphics[width=.9\textwidth]{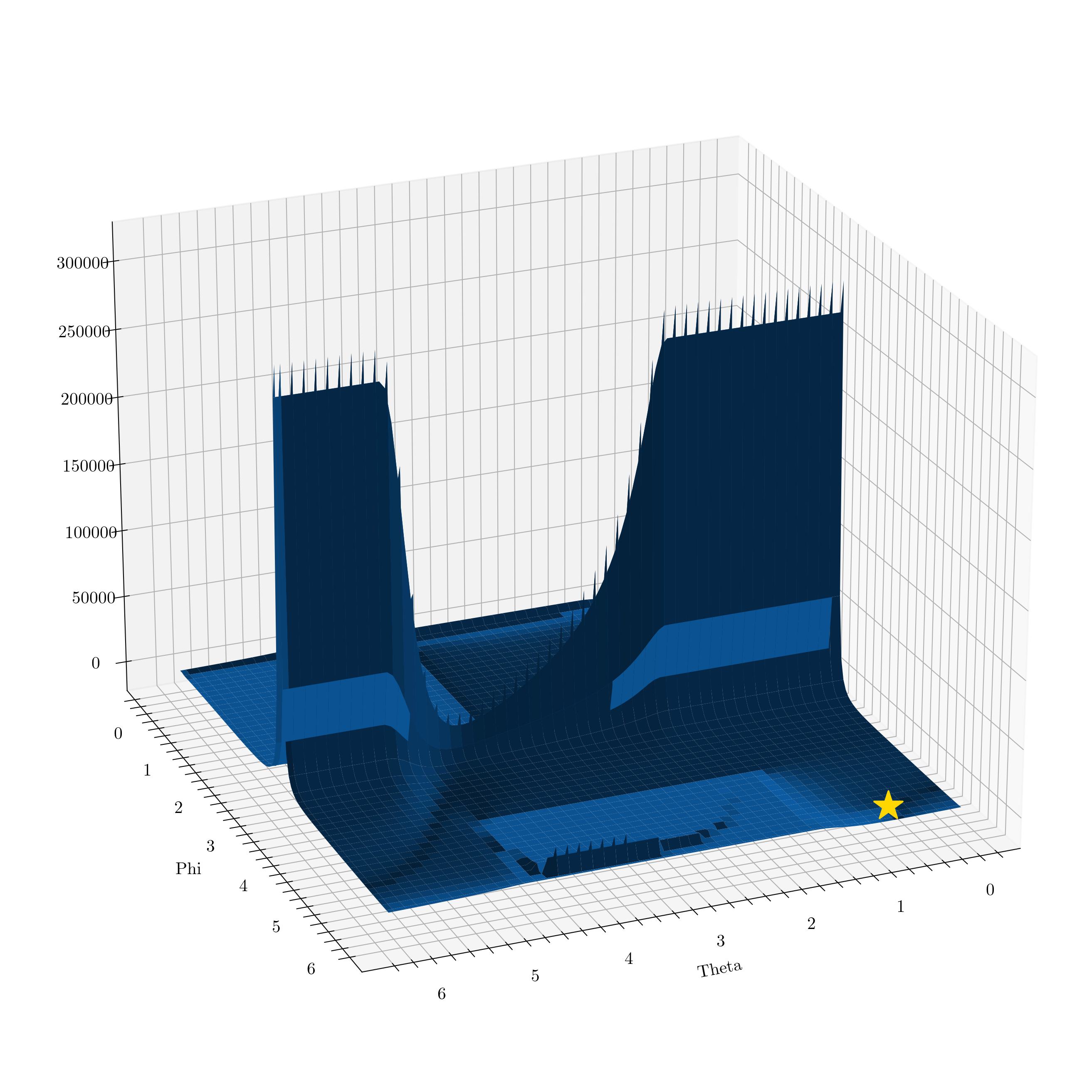}
    \end{subfigure}%
    \begin{subfigure}{.5\textwidth}
        \centering
        \includegraphics[width=.9\textwidth]{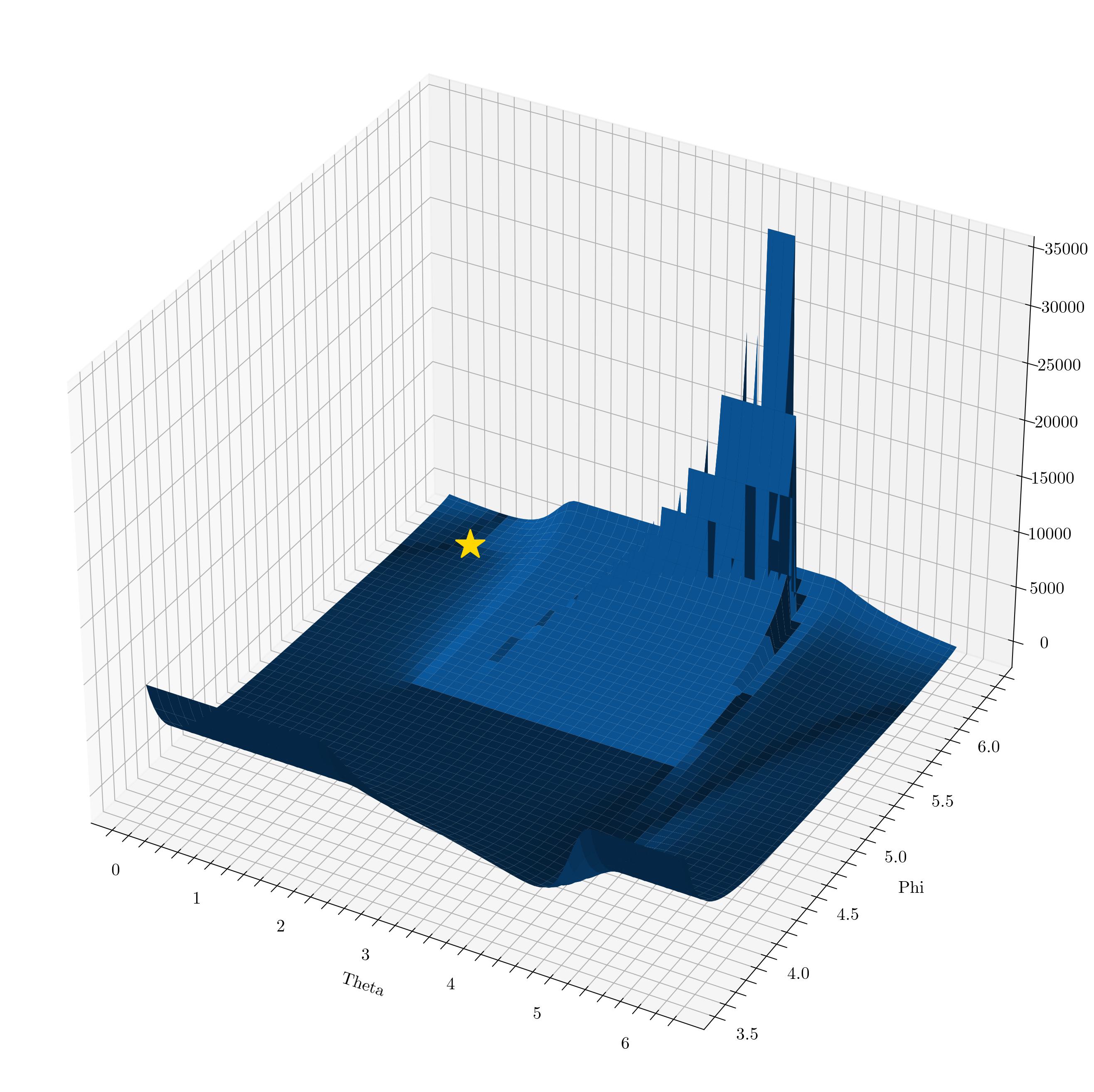}
    \end{subfigure}
    \caption{Two views of the loss function for patient Normal1 with different bounds. The right plot has bounds $[0, 2\pi] \times [0, 2\pi]$ while the left plot has bounds $[0, 2\pi] \times [4, 2\pi]$. The gold star represents the point of optimal loss for the bounded loss function. Observe that there are plotting artifacts due to very high gradients.}
    \label{fig:unboundedloss}
\end{figure}

\begin{figure}[H]
    \centering
    \includegraphics[width=.6\textwidth]{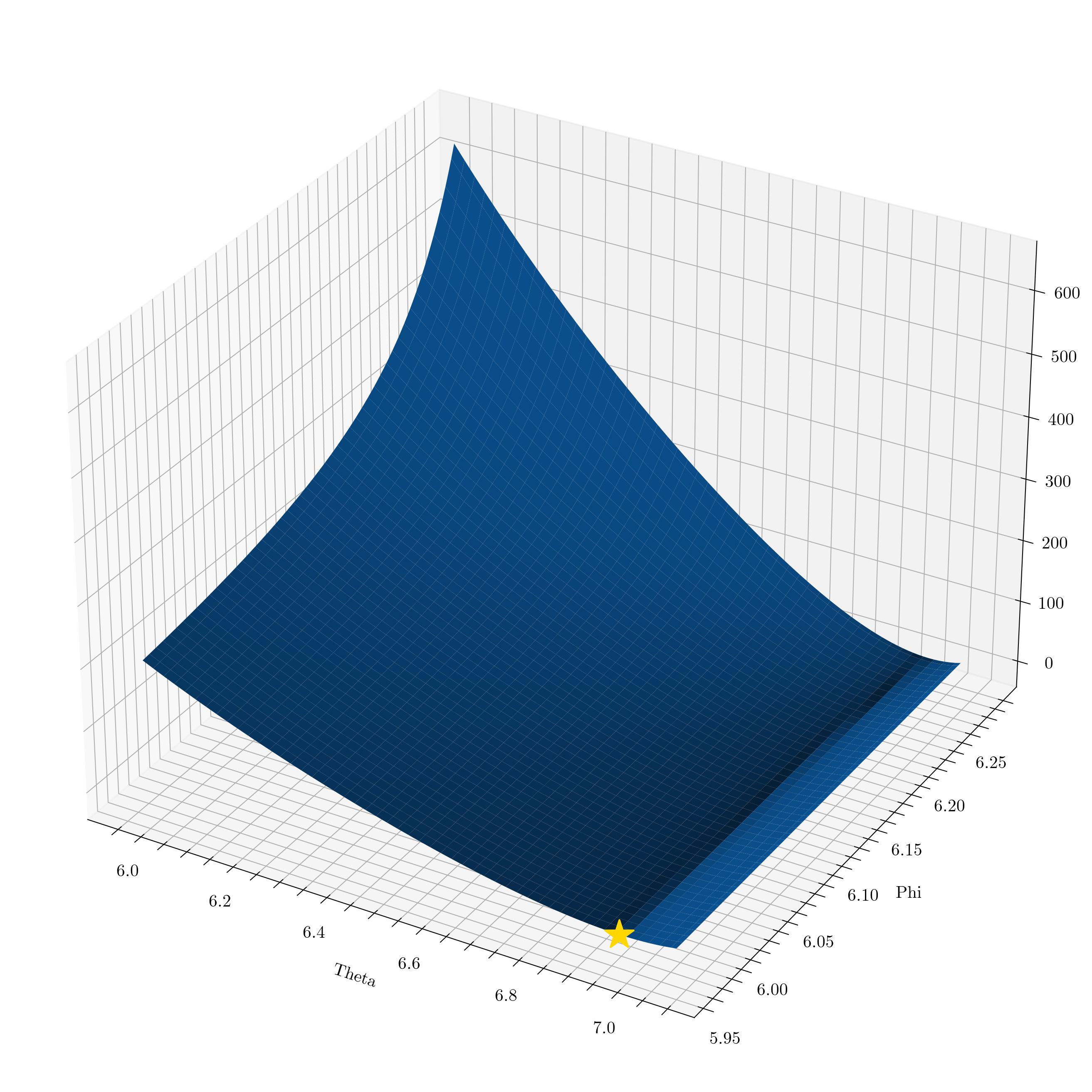}
    \caption{The loss function for patient Normal1, plotted as a surface in three dimensions and bounded via the scheme described in the Application to Dynamic FDG-PET section. The $x$ and $y$ axes are for the input parameters of the loss function (in radians), while the $z$ axis is for the unit-less loss. The gold star represents the point of optimal loss on this loss function.}
    \label{fig:boundedloss}
\end{figure}

% More data needed here--unbounded loss function and up-to-date loss functions
Since our optimization is two dimensional, we can directly visualize the loss function for our optimization problem to answer this question. We found that the loss function was generally poorly behaved when we did not bound the loss function. In particular, we saw very high gradients and possible locations of non-convexity. Figure \ref{fig:unboundedloss} shows the loss function for patient Normal1 without bounds (left) and with liberal bounds (right). However, applying the bounding scheme mentioned in the Application to Dynamic FDG-PET section, we found that the loss function was relatively well-behaved. Figure \ref{fig:boundedloss} shows the bounded loss function for Normal1. The loss function is overall smooth and convex. At the point of optimal loss, however, the gradient in the direction of $\phi$ is small---possibly indicating poor estimation of $\phi$. This is indeed reflected by the results in \textbf{Q4}.

\subsection{Q4: Does the estimated blood input match well with the ground truth blood input?}
\begin{figure}[H]
    \centering
    \includegraphics[width=\textwidth]{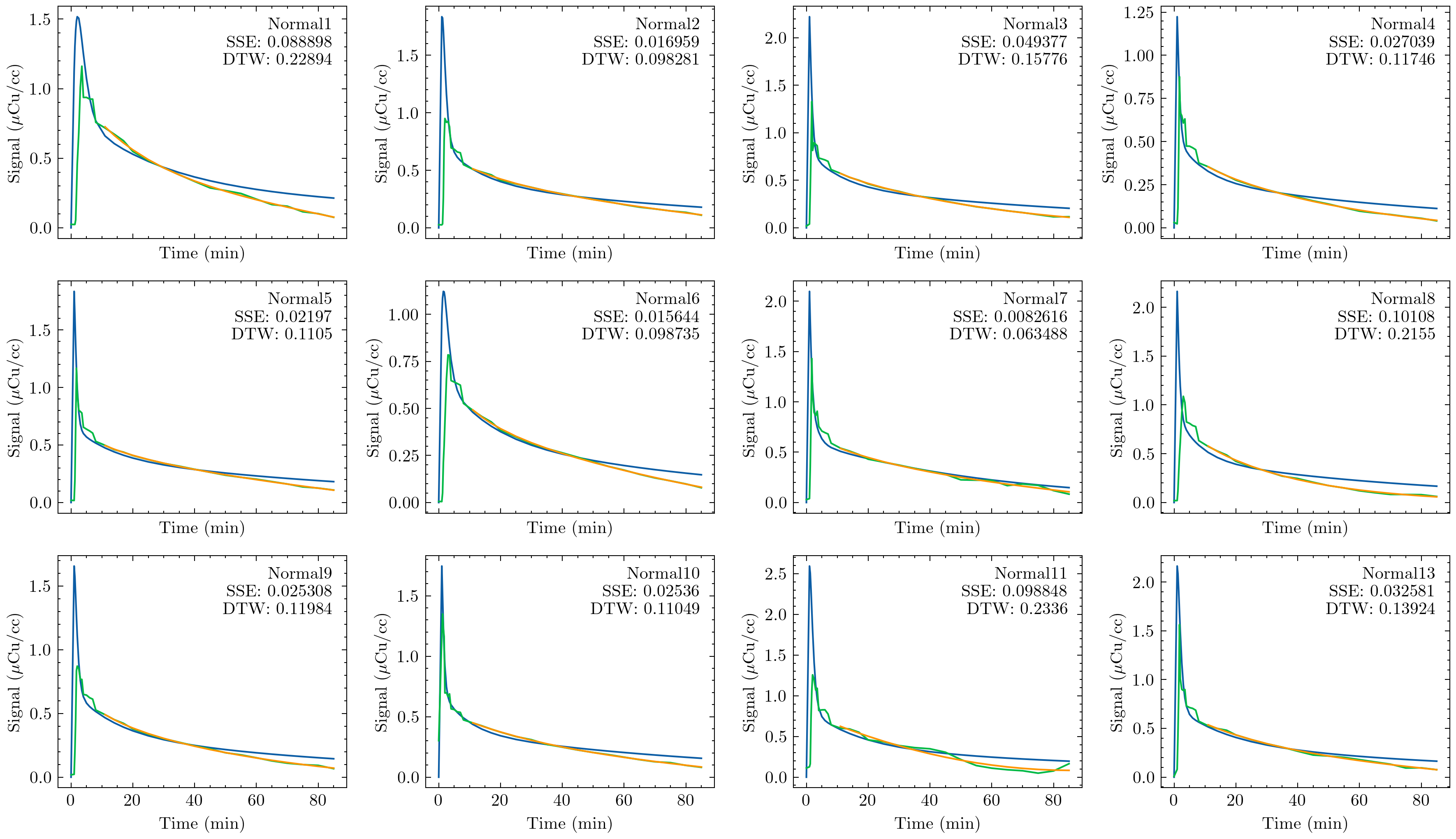}
    \caption{The estimated blood input function (yellow) scaled to best match the ground truth blood input versus the ground truth blood input (blue). A full-time blood input is also extrapolated (green) by applying the same linear combination used to obtain the partial blood input function to the full-time tissue activity principal components.}
    \label{fig:est_vs_blood_bestfit}
\end{figure}

\begin{figure}[H]
    \centering
    \includegraphics[width=\textwidth]{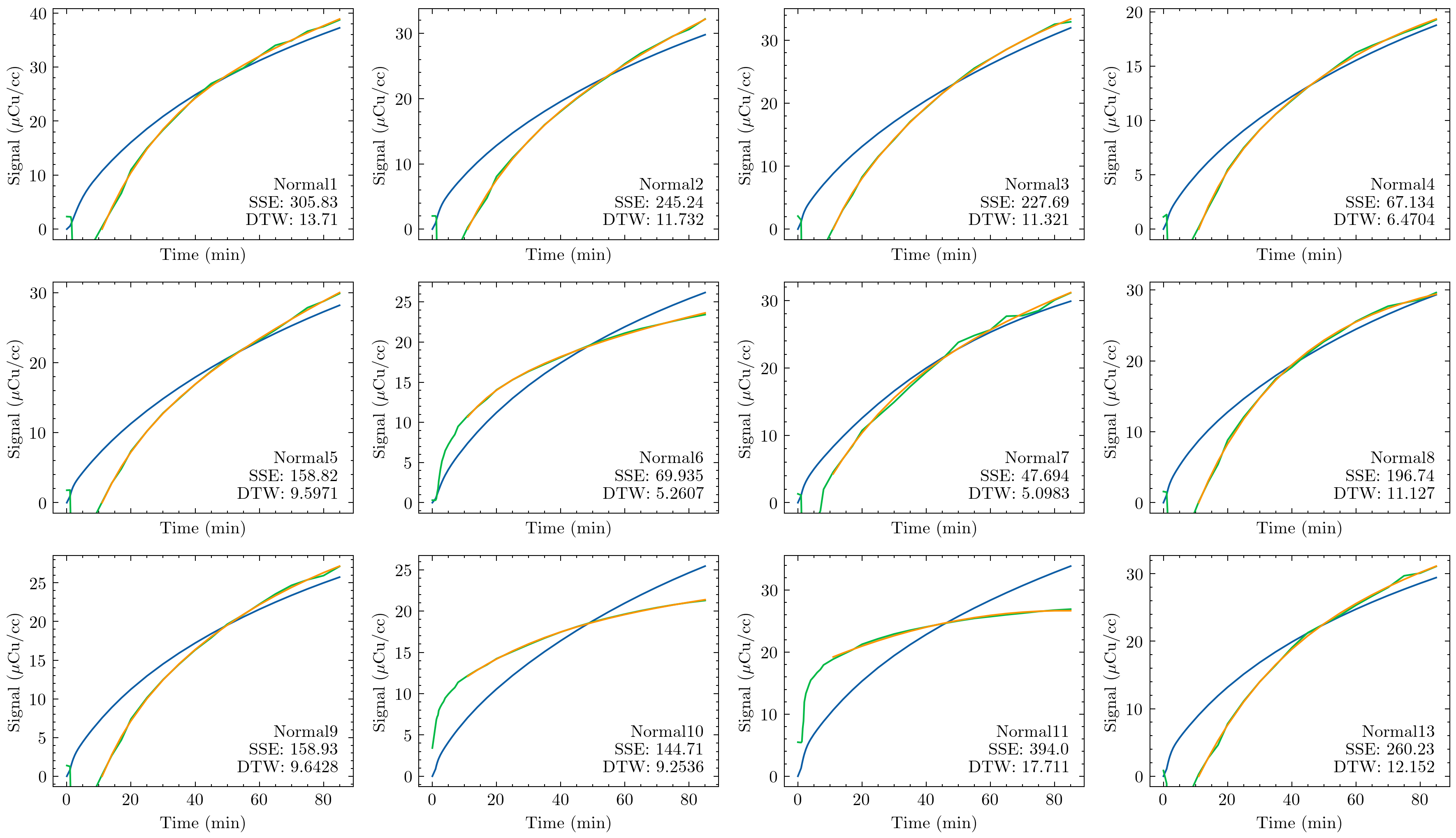}
    \caption{The estimated integral of the blood input function (yellow) scaled to best match the ground truth versus the ground truth (blue). The full-time estimated integral of blood is extrapolated (green) similar to Figure \ref{fig:est_vs_blood_bestfit}. Note the very poor correspondence in line with our observations of the loss function.}
    \label{fig:est_vs_int}
\end{figure}

% obtain results for scale by endpoint
After scanning our patients and simultaneously obtaining physically drawn blood input functions, we applied our methods. Figures \ref{fig:est_vs_blood_bestfit} and \ref{fig:est_vs_int} show our results. Note that our method produces a blood input function without scale, so we scale in both figures by the factor which gives the best correspondence between the estimate and ground truth. In practice, one's scale factor would be less precise. 

Ultimately, we obtained an average sum of squared errors of $0.042611 \pm  0.032533$ and dynamic time warping distance of $0.14115 \pm  0.053716$ between the blood input function and the best-scaled estimated blood input function. These results overall show good correspondence, though the estimated blood input function is consistently flatter at the tail than the ground truth blood input function. We obtained an average sum of squared errors of $189.75 \pm  99.26$ and dynamic time warping distance of $10.256 \pm  3.4504$ between the ground truth integral of blood and estimated, best-scaled integral of blood. As with \textbf{Q3}, this shows poor estimation of $\phi$, which determines the estimated integral of the blood input function.


\subsection{Q5: Do the estimated and ground truth $K_i$-maps match well?}
% patient-by-patient labels, change z-score differenece to absolute difference? must be consistent
After obtaining the estimated blood input functions, we generated voxel-wise Patlak $K_i$-maps for each of our scans based on the estimated and true blood input functions. Figure \ref{fig:img} shows our results. Recall that our method gives us a blood input function without scale. Unlike \textbf{Q4} where we scaled by the factor which best matched the estimated and true blood input functions, we instead normalize both the estimated and ground truth $K_i$-maps in Figure \ref{fig:img}.

Table \ref{tab:img} shows our numerical results. We ultimately got an average root mean squared error (RMSE) of $8.8636\mathrm{e}{-4} \pm  2.6702\mathrm{e}{-4}$, an average mean absolute difference between the z-scores of $0.036827 \pm  0.015544$, an average mean absolute percentage error (MAPE) of $2.5394 \pm  0.92892$, and an average structural similarity metric (SSIM) of $0.99148 \pm  0.0054024$. Note that for the RMSE, MAPE, and mean z-score difference the means are taken over segmentation of the head defined by thresholding the scan. Additionally, unlike the figure, the RMSE is between the unnormalized ground truth $K_i$ map and the estimated $K_i$ map, scaled so that the means of the two images agree. Overall, these results show good correspondence between the true and estimated $K_i$ maps.

\begin{figure}[H]
    \centering
    \includegraphics[trim={0 25cm 0 0}, clip, width=\textwidth]{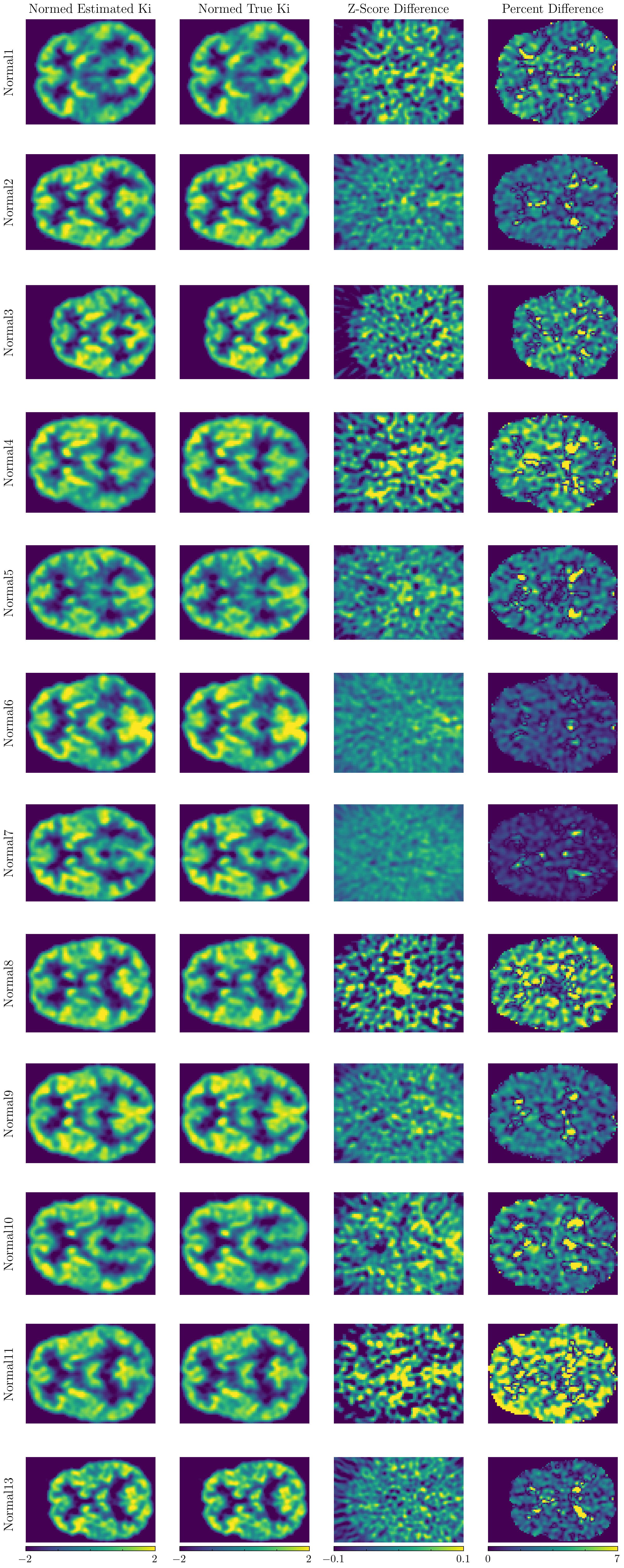}
    \phantomcaption
\end{figure}
\begin{figure}[H]
    \ContinuedFloat
    \centering
    \includegraphics[trim={0 0 0 26cm }, clip, width=\textwidth]{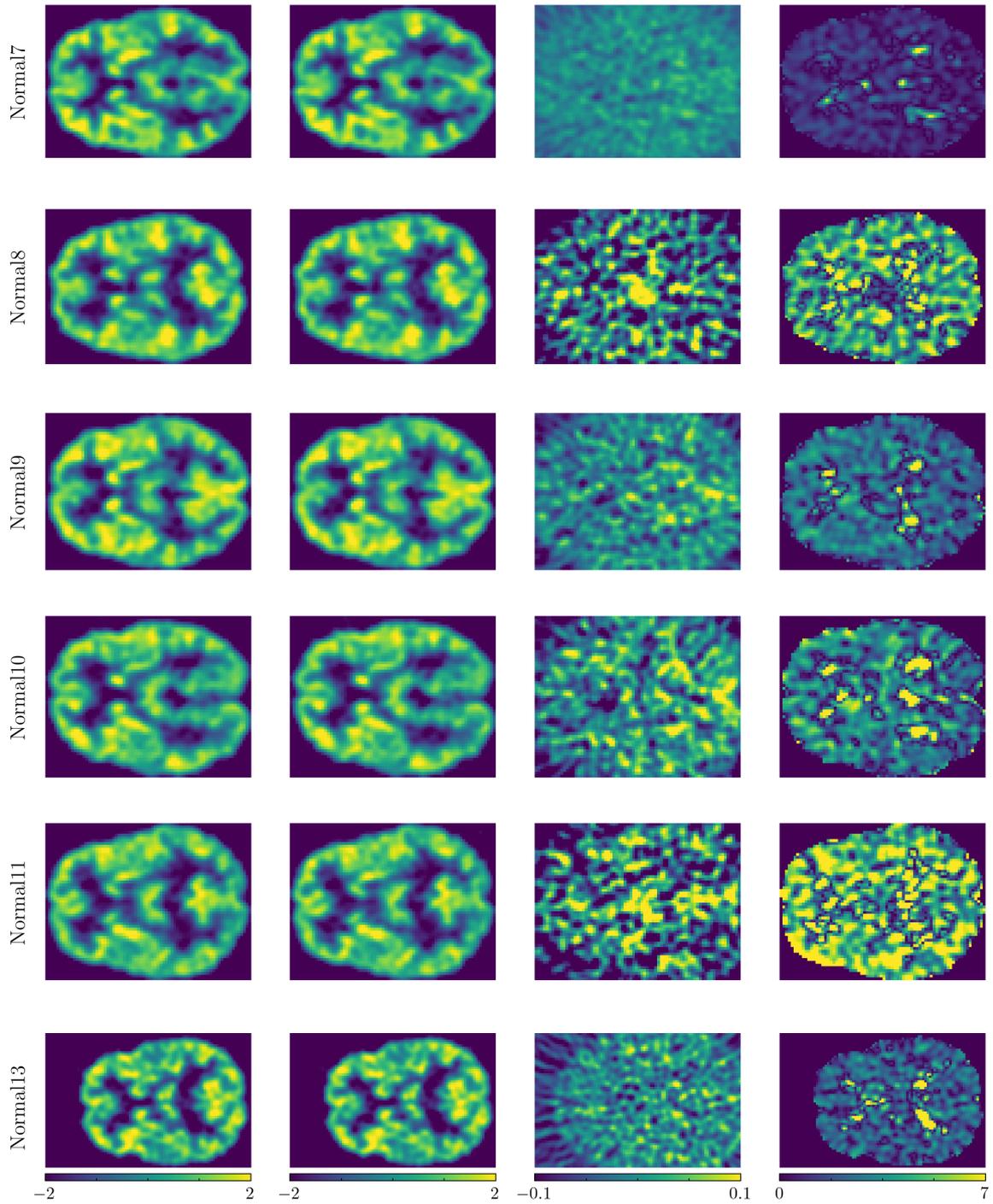}
    \caption{One coronal slice of the voxel-wise estimated and true Patlak maps for each patient. Columns are, from left-to-right, the estimated $K_i$-map normalized to mean of $0$ and standard deviation of $1$, the true $K_i$-map normalized similarly, the normalized estimated $K_i$ map minus the ground truth $K_i$ map, and the mean absolute percent difference between true and estimated.}
    \label{fig:img}
\end{figure}

% Absolute z-diff:  0.036827 \pm  0.015544
% Absolute percent diff:  2.5394 \pm  0.92892
% Absolute ssim:  0.99148 \pm  0.0054024
% Absolute rmse:  0.00088636 \pm  0.00026702
\begin{table}[H]
    \centering
    \begin{tabular}{lcccc}
        \toprule
        Patients&Mean Z-Diff&RMSE&MAPE&SSIM\\
        \midrule
        Normal1&0.0423&9.25E-04&2.83&0.991 \\
        Normal2&0.0259&7.17E-04&2.03&0.996 \\
        Normal3&0.0371&1.04E-03&2.96&0.989 \\
        Normal4&0.0457&1.29E-03&3.21&0.987 \\ 
        Normal5&0.0328&7.40E-04&2.50&0.993 \\
        Normal6&0.0172&6.92E-04&1.31&0.998 \\
        Normal7&0.0136&3.83E-04&0.876&0.999 \\
        Normal8&0.0614&1.20E-03&3.62&0.983 \\
        Normal9&0.0281&8.47E-04&2.10&0.995 \\
        Normal10&0.0399&8.30E-04&2.75&0.991 \\
        Normal11&0.0683&1.30E-03&4.36&0.981 \\
        Normal13&0.0296&6.65E-04&1.93&0.995 \\
        \hline
    \end{tabular}
    \caption{Individual similarity scores for each patient. The Mean Z-Diff column represents the mean absolute difference between the z-scores for the predicted and ground truth $K_i$ maps. The RMSE, MAPE, and SSIM columns respectively stand for the root mean squared error, the mean absolute percentage error, and the structural similiarity metric.}
    \label{tab:img}
\end{table}

\section{Discussion}

Dynamic PET measures radiation over a series of time windows for about 60 minutes; the conventional analysis performs convolution of the 4D brain PET data with the blood input in a dual output kinetic model to form a 3D parametric brain PET map, which according to our pilot studies, provides meaningful information not available from standard static PET\cite{seshadri_dynamic_2021, quigg_dynamic_2022, hossain_multimodal_2023, kundu_dynamic_2024, kundu_interictal_2024}. Clinical adoption is still a challenge however due to difficult analysis protocols, especially computation of the blood input function which evaluates the amount of tracer in the blood the tissue can use. The gold standard for measuring this function is through time-distributed arterial blood sampling during the scan, which is costly and risks patient infection or arterial occlusion. There have been several developments over the years in the derivation of image-derived blood input function (IDIF) which for rodents could be from the left ventricular cavity or the inferior vena cava or the internal carotid arteries for total body images\cite{li_improved_2018, huang_non-invasive_2019, massey_model_2021}. Deriving IDIF from human images is challenging especially for brain PET limited field of view image data. These data require segmenting the internal carotid arteries for IDIF derivation which could be inefficient and user dependent. A prior work compared three different methods including local means analysis, soft-decision similar component analysis, and k-means for automated internal carotid artery (ICA) segmentation for human dynamic FDG brain PET studies\cite{zanotti-fregonara_comparison_2009}. Recent works utilize MRI\cite{traub-weidinger_utility_2020} and CT\cite{shah_automatic_nodate} based land-marking techniques for blood input determination from the heart for whole-body dynamic PET images using continuous bed motion, which may be a limitation for several imaging centers. New work from our lab developed a deep learning pipeline for automated frame selection and segmentation of the ICA for IDIF derivation for limited field of view dynamic FDG brain PET data\cite{chavan_end--end_2024}. The model training however requires creating ground truth labels for ICA frame selection and segmentation. Recent new work developed an automatic method combining wavelets and unsupervised learning for isolating arterial IDIF which identifies curves that have large peaks and small tails\cite{moradi_automated_2024}. The authors compare their methods with ground truth image-derived IDIF generated from the descending aorta without accounting for partial volume averaging and cardiac and respiratory motion. Spill over and partial volume effects would severely confound ground truth IDIF and hence validity of the computed arterial curves. 

This work for the first time develops a mathematical formalism for automatic derivation of IDIF using the graphical patlak model and principal component analysis. Our method automatically derived blood input and donwstream whole brain FDG uptake rate maps, which compared well with ground truth arterial blood samples and $K_i$ maps for dynamic FDG brain PET data in control subjects. The study is however not without limitations. The model assumes that the Patlak model holds after a certain time point when the tracer uptake equilibrates in the blood and tissue. This along with the parametric formulation of the blood input may be a limitation for tracers which exhibit reversible behavior unlike FDG.

%\subsection{Limitations}
%\textbf{Limitations:} The study is not without limitations.

\section{Conclusion}
Although not often viewed as such, the assumption that some dynamic data follow a kinetic model gives strong prior information. In the case of the Patlak plot, we can use this assumption to estimate an unscaled blood input function and an unscaled $K_i$ map.

\section*{Acknowledgments}
WT and BK was supported in part by grants from the University of Virginia Brain Institute, Ivy innovation funds, Commonwealth commercialization funds (CCF) from Virginia Innovation Partnership Corporation and start-up funds from the Department of Radiology and Medical Imaging at the University of Virginia (all to BK). BK was also supported in part by NIH grants R01EB029450 (PI, Berr, SS) and R01EB030744 (PI, Price, R). MM was supported in part by grants from the NIH grant number R50-CA211270 (PI, Muzi, M). Imaging and blood sampling data were collected under a program project grant, P01-CA042045 (PI,Krohn, KA).  All patients signed informed consent and the project was approved by the local IRB (Institutional Review Board), Radiation and Human Subject committees at the University of Washington, Seattle.

\bibliographystyle{unsrt}
\bibliography{Radiology}
\end{document}